
\magnification=\magstep1
\overfullrule=0pt
\hsize=15.4truecm
{\nopagenumbers \line{\hfil UQAM-PHE-95/02}
\vskip2cm
\centerline{\bf $K^+\rightarrow \pi^+\nu\overline\nu$\
AND $K^0_L\rightarrow\mu^+\mu^-$\ DECAYS WITHIN THE}
\centerline{\bf MINIMAL SUPERSYMMETRIC STANDARD MODEL}
\vskip2cm
\centerline{G. COUTURE AND H. K\"ONIG
\footnote*{email:couture, konig@osiris.phy.uqam.ca}}
\centerline{D\'epartement de Physique}
\centerline{Universit\'e du Qu\'ebec \`a Montr\'eal}
\centerline{C.P. 8888, Succ. Centre Ville, Montr\'eal}
\centerline{Qu\'ebec, Canada H3C 3P8}
\vskip2cm
\centerline{\bf ABSTRACT}\vskip.2cm\indent
We present a detailed calculation of the contributions of charginos, scalar
quarks, and charged Higgs boson
to the $K^+\rightarrow \pi^+\nu\overline\nu$\
and $K^0_L\rightarrow\mu^+\mu^-$\ decays.
We include mixings: that of charginos and that of the
scalar partners of the left and right handed top quark.
We find that the box contribution to the amplitudes
is much smaller than the penguin contribution, which can be $\sim$10\% of the
Standard Model contribution, even for relatively large SUSY masses.
The charged Higgs contribution can be as large as 25\%
of the SM contribution in the first decay and as much as 40\% of
the SM contribution in the second decay.
\vskip1cm
\centerline{ March 1995}
\vfill\break}
\pageno=1
\noindent
{\bf I. INTRODUCTION}\hfill\break\vskip.2cm
\noindent
Rare decays have always been a good field to search for new physics. Among
those, rare Kaon decays have been longtime favorites [1]. As the mass of the
the top quark increases, within the SM those rare decays become essentially
{\it
top physics}. In this note, we will consider two closely related
processes: $K_L^0\to\mu^+\mu^-$ and $K^+\to\pi^+\nu\bar\nu$. The first
branching ratio is
now measured with a precision of a few percent [2]: $7.4\pm 0.4\times 10^{-9}$.
The upper bound on the second one ($5.2\times 10^{-9}$) is now getting close to
the predicted value in the SM: $\sim 6\times 10^{-10}$ [1,3-7] for large
top quark mass. The E787 group at Brookhaven National Laboratory aims to
measure this branching ratio in the near future.
Once the mass of the top quark is measured with good precision,
these rare decays will open a window on physics
beyond the SM. One of the favorite such models is the Minimal
Supersymmetric Standard Model (MSSM)[8].
The effects of the MSSM to the decays mentioned
above were considered in many papers a while ago [9--13].
However none of them included the full mixing of the
charginos and the scalar partner of the left and right
handed top quark. In this paper we present a more
complete calculation of the SUSY contributions to the
$K^+\rightarrow\pi^+\nu\overline \nu$\ and
$K_L^0\rightarrow\mu^+\mu^-$\ within the MSSM.
We include the mixing of the charginos as well as
the mixing of the scalar partner of the left and right
handed top quark, which is proportional to the top
quark mass. This last mixing was rightfully neglected in the past but now that
the top quark mass of getting very large, one should not neglect it.
We neglect all masses of the fermions
compared to the SUSY masses except the mass of the
top quark, which we will take to be the recently
released CDF value of 174 GeV [14].
\hfill\break\vskip.2cm\noindent
{\bf II. SUSY CORRECTIONS}\hfill\break\vskip.2cm\noindent
\line {\bf Constraints and mixing\hfil}
\noindent
Before we proceed with the calculation of the decay amplitudes, we should
discuss briefly some limits on two important parameters of SUSY; namely
$\mu$ and $m_{g_2}$.
 These symmetry-breaking parameters are independant but since
they are
used to generate masses for particles, we can put constraints on them from
current experimental bounds on SUSY masses. We will use here the masses of the
charginos ($\tilde W_i$). Following [8] we define a mixing matrix
$$
X = \left(\matrix{m_{g_2} & m_W\sqrt{2}\sin\beta \cr
                  m_W\sqrt{2}\cos\beta& \mu \cr}\right)\eqno(1)
$$
\noindent
where $\tan\beta = v_2/v_1 $\ is the ratio of the vacuum expectation
values (vev's) of the scalar Higgs bosons in the MSSM.\footnote
*{Note that $\tan\beta=v_1/v_2$\ in [8].}
One then proceeds to diagonalize this matrix
through unitary matrices such that $U^*XV^{-1} = M_D$.
One obtains two
eigenvalues for the masses:
$$\eqalign{
m_{\tilde W_{1,2}}^2 =
& {1\over 2}\Big\lbrace m_{g_2}^2+\mu^2+2m_W^2\hfil\hfil\cr
           &\pm\sqrt{(m_{g_2}^2-\mu^2)^2+4m_W^4\cos^22\beta
 +4m_W^2(m_{g_2}^2+\mu^2+2m_{g_2}\mu\sin 2\beta)}\Big\rbrace
}\eqno(2)
$$
and the unitary matrices take the form
$U= O_-$ and $V=O_+$ if $\det X\ge 0$ while $V=\sigma_3O_+$ if $\det X< 0$
with
$$
O_\pm = \left(\matrix{\cos\phi_\pm& \sin\phi_\pm\cr
		    -\sin\phi_\pm& \cos\phi_\pm\cr}\right)\eqno(3)
$$
\noindent
with
$$
\eqalign{
\tan2\phi_-& = 2\sqrt{2}m_W(\mu~ sin\beta +
m_{g_2} \cos\beta)/(m_{g_2}^2-\mu^2-2m_W^2
\cos2\beta)\cr
\tan2\phi_+& = 2\sqrt{2}m_W(\mu
\cos\beta + m_{g_2} \sin\beta)/(m_{g_2}^2-\mu^2+2m_W^2
\cos2\beta)}\eqno(4)
$$
\noindent
In order to keep $M_D^{11} \leftrightarrow M_+ > M_- \leftrightarrow M_D^{22}$
one must impose $0\le \phi_\pm\le \pi/2$.
\noindent
One gets a relationship between $m_{g_2}$ and $\mu$ through the previous roots:
we require these to be larger than 50 GeV since charginos
($\tilde W_i$) have
not been observed. We use $m_W = 80.1~GeV$. This gives us Fig.1 for
different values of $\tan\beta$.
The regions above the upper curves and below
the lower curves are allowed; {\it ie} both roots will be larger than 50 GeV.

We also include the mixing in the scalar top quark sector. The mass matrix we
are dealing with has the form
$$
M_{\tilde t}^2 =
\left(\matrix{m_{\tilde t_L}^2+m_{\rm top}^2+0.35~D_Z&
 -m_{\rm top}(A_{\rm top}+\mu\cot\beta)
\cr
-m_{\rm top}(A_{\rm top}+\mu\cot\beta)&
m_{\tilde t_R}^2+m_{\rm top}^2+0.15~D_Z\cr}\right)
\eqno(5)$$
\noindent
where $D_Z = m_Z^2\cos2\beta$ and $A_{\rm top}$
is a parameter that describes the
strenght of nonsupersymmetric trilinear scalar interactions; we set
$A_{\rm top} = m_S$, where $m_S$\ is the soft SUSY breaking mass term.
We also take $m_{\tilde t_L}$ and $m_{\tilde t_R}$ equal to $m_S$.
Note also that $0.35 = T_3^{\rm top} - e_{\rm top}\sin^2\theta_W$ and
$0.15 = e_{\rm top}\sin^2\theta_W$.
Instead of working directly with the current eigenstates $\tilde t_{L,R}$\ we
work with the mass eigenstates
$$\tilde t_1=cos\Theta\tilde t_L+\sin\Theta\tilde t_R\qquad
  \tilde t_2=-\sin\Theta\tilde t_L+\cos\Theta\tilde t_R\eqno(6)
$$
\noindent
The different components are obtained by diagonalisation of the previous
matrix:
$$
tan2\Theta={2m_{\rm top}(A_{\rm top}+\mu\cot\beta\over
m_{\tilde t_R}^2-m_{\tilde t_L}^2-0.20~D_Z}
$$
\noindent
The two mass eigenstates $m_{\tilde t_{1,2}}$\ are given by
$$
m_{\rm top}^2+{1\over 2}\left( m_{\tilde t_L}^2
+m_{\tilde t_R}^2+0.5 D_Z\pm\sqrt{
(m_{\tilde t_R}^2-m_{\tilde t_L}^2-0.20D_Z)^2+
4m_{\rm top}^2(A_{\rm top}+\mu\cot\beta)^2}\right)\eqno(7)
$$
\noindent
In order to recover the proper masses when there is
no mixing, one has to select $m_{\tilde t _1}^2 \to$
 negative root and
$m_{\tilde t_2}^2\to$
positive root. Again, we constrain the angle $\Theta$ to the first
quadrant.
For a more complete description of the mass matrix for the
scalar top quarks and the couplings of the Z boson to
the scalar top quarks within the scalar top quark mass
eigenstates we refer the reader to the literature [16,17]

We now proceed to the loop calculations themselves.
The decay modes occur via box and penguin diagrams.
In the SM it turns out that the contribution of
the box diagram is negligible compared to the
contribution of the penguin diagram for a large
top quark mass. In the SM the amplitude of the decay
$K_L^0\rightarrow\mu^+\mu^-$\ via the box
diagram is suppressed by a relative factor of 4
compared to the decay $K^+\rightarrow\pi^+\nu\overline \nu$\
[3], whereas for the penguin diagram
there is a relative minus sign and the function
$D(y_j,x_t)$\ (eq.2.15 in [5]) has to be replaced by $C(x_t)$\
(eq.2.14 in [5]).
\hfill\break\indent
In the MSSM these decays proceed
through the box diagrams in Fig.2 and
the penguin diagrams in Fig.3. In Fig.2 we also include
the so called mass insertion diagram. In [12] the authors
only considered the first diagram in Fig. 2 whereas the
authors in [9] put their emphasis on the second one.
We include both and take the full coupling as given
in Fig.22 and Fig.23 of [15]\footnote*{In Fig.22 b)+d)
and Fig.23 b)+d) $\gamma_5$\ has to be replaced
by $-\gamma_5$}.
\hfill\break
\line{\bf Box Diagrams\hfil}
\noindent
To calculate the box diagrams in Fig.2, we have used
the rules given in appendix D of Ref. 8.
After a lenghty calculation, the result\footnote\dag{All our results are given
for one neutrino family.} from the box diagram
for the decay $K^+\rightarrow\pi^+\nu\overline \nu$\
is given by:
$$\eqalignno{iM_{BOX}=&{{\alpha^2}\over{4\sin\theta_W^4}}
K_{td}K^*_{ts}(V_{\tilde t}-V_{\tilde u})\overline u_\nu
\gamma_\mu P_L v_{\overline\nu}\overline v_s\gamma^\mu
P_L u_d&(8)\cr
V_{\tilde u}=&\sum\limits_{i,j=1,2}U_{i1}U_{j1}V_{i1}
V_{j1}\lbrack \tilde F^{ij}_{\tilde l\tilde u}
+2 _M\tilde F^{ij}_{\tilde l\tilde u}\rbrack\cr
V_{\tilde t}=&\sum\limits_{i,j=1,2}U_{i1}U_{j1}
\Bigl\lbrace\bigl\lbrack V_{i1}V_{j1} c^2_\Theta+
V_{i2}V_{j2}{{m^2_{\rm top}}\over{2m_W^2\sin^2\beta}}
s^2_\Theta\bigr\rbrack\lbrack\tilde F^{ij}_{\tilde l
\tilde t_1}+2 _M\tilde F^{ij}_{\tilde l\tilde t_1}\rbrack\cr
&+\bigl\lbrack V_{i1}V_{j1} s^2_\Theta+V_{i2}V_{j2}
{{m^2_{\rm top}}\over{2m_W^2\sin^2\beta}}
c^2_\Theta \bigr\rbrack\lbrack F^{ij}_{\tilde l\tilde t_2}
+2 _M\tilde F^{ij}_{\tilde l\tilde t_2}\rbrack\Bigr\rbrace
\cr}$$
$c^2_\Theta=\cos^2\Theta$, $s^2_\Theta=\sin^2\Theta$.
$\tilde F^{ij}_{\tilde l
\tilde u, \tilde t_{1,2}}$\  and $_M\tilde F^{ij}_{\tilde l
\tilde u, \tilde t_{1,2}}$\ are given in the appendix A.
$m_{i,j}=m_{\tilde W_{i,j}}$\ are the mass eigenvalues
of the charginos, $m_{\tilde l}$\ the mass of the
scalar leptons and $m_{\tilde u, \tilde t_{1,2}}$\
the masses of the scalar up quark and the eigenstates
of the scalar top quark including the mixing.
$V_{ij}$\ and $U_{ij}$\ are the diagonalizing
matrices of the charginos as given in eq. C19 in [8]
taken to be real. One can show that
the result of the box diagram for the decay
$K_L^0\rightarrow\mu^+\mu^-$\ can be obtained from these results simply
by replacing $U_{i1}U_{j1}$\ with $-V_{i1}V_{j1}$ and $m_{\tilde l}$\
with $m_{\tilde\nu}$.
\hfill\break\indent
In the case with no mixing of the charginos
and scalar top quark and when neglecting the
mass insertion term and dropping the $m^2_{\rm top}$\
terms eq.(8) agrees with eq.(1)
in [12] up to a factor of 2.\footnote *{Interchanging of
$\mu^+(\overline\nu)$\ with $\mu^-(\nu)$\ leads to the
mass insertion term diagrams, which give a different
function and thus not only a factor of 2.}
Note also that $\tilde F^{ii}_{\tilde l
\tilde u}\equiv -g(\tilde y_j,\tilde y_u)/\tilde m_W^2$\
in [12]. Whereas the function $g$\ in eq.(2) in [12]
is described as the function $g_1(x,y)$\ in eq.C.2
in [5] we have that the function $_M\tilde F^{ii}
_{\tilde l\tilde u}$\ is described by the function
$g_0(x,y)$\ in eq. C.2 in [5].
\hfill\break\indent
The box diagram with the charged Higgs boson and the
up quarks is proportional to the lepton masses and therefore
negligible.

As it turned out, the box contribution was always much smaller than the
penguin contribution (except when the penguin was 0, by {\it numerical
accident}) and totally negligible.
\hfill\break
\line{\bf Penguin contributions\hfil}
\noindent
When considering penguin diagrams
involving the strong coupling constant
one usually first
considers the diagrams with gluinos and scalar
down quarks within the loop. After a lengthy but
straightforward calculation one can show that the
$\overline v_s\gamma_\mu P_L u_d$\ is identical
to 0 after Feynman integration and one is left
with terms proportional to
$(q_\mu q_\nu-q^2g_{\mu\nu})\overline v_s\gamma^\nu P_L u_d$\
and $i\sigma^{\mu\nu}q_\nu\overline v_s(m_sP_L+m_dP_R)u_d$,
which since $q^2\ll m_Z^2$\ and $m_{s,d}\ll m_Z$\ are
totally negligible [18]. This result was also obtained
in [12].
 The same argument goes when neutralinos
and scalar down quarks are taken within the loop.
The result for the penguin diagrams in Fig.3
with charginos and scalar quarks within the loop is given by:
$$\eqalignno{iM_{\rm Peng.}=&+{{g_2^3}\over{(4\pi)^2
\cos\Theta_W}}K_{td}K^*_{ts}(-M_{2\tilde q}+M_{2\tilde\chi}
+M_{SE})\overline v_s\gamma_\mu P_Lu_d&(9)\cr
M_{2\tilde q}=&\sum\limits_{i=1,2}\Bigl\lbrace V^2_{i1}
\bigl\lbrack (T_{3t}c^2_\Theta-e_ts_W^2)c^2_\Theta
\tilde T^{\tilde t_1\tilde t_1}_i+(T_{3t}s^2_\Theta-
e_ts_W^2)s^2_\Theta\tilde T^{\tilde t_2\tilde t_2}_i\cr
&+2c^2_\Theta s^2_\Theta T_{3t}\tilde T^{\tilde t_1
\tilde t_2}_i\bigr\rbrack\cr
&+{{m^2_{\rm top}}\over{2m_W^2\sin^2\beta}}V^2_{i2}
\bigl\lbrack (T_{3t}c^2_\Theta-e_ts_W^2)s^2_\Theta
\tilde T^{\tilde t_1\tilde t_1}_i+(T_{3t}s^2_\Theta-
e_ts_W^2)c^2_\Theta\tilde T^{\tilde t_2\tilde t_2}_i\cr
&-2c^2_\Theta s^2_\Theta T_{3t}\tilde T^{\tilde t_1
\tilde t_2}_i\bigr\rbrack\cr
&-{{m_{\rm top}}\over{\sqrt{2}m_W\sin\beta}}V_{i1}V_{i2}
2s_\Theta c_\Theta
\bigl\lbrack (T_{3t}c^2_\Theta-e_ts_W^2)
\tilde T^{\tilde t_1\tilde t_1}_i-(T_{3t}s^2_\Theta-
e_ts_W^2)\tilde T^{\tilde t_2\tilde t_2}_i\cr
&-(c^2_\Theta- s^2_\Theta) T_{3t}\tilde T^{\tilde t_1
\tilde t_2}_i\bigr\rbrack\Bigr\rbrace\cr
M_{2\tilde\chi}=&\sum\limits_{i,j=1,2}\Bigl\lbrace
V_{i1}V_{j1}\lbrack c^2_\Theta\tilde G_1^{ij}+
s^2_\Theta\tilde G_2^{ij}\rbrack+
{{m^2_{\rm top}}\over{2m_W^2\sin^2\beta}}V_{i2}V_{j2}
\lbrack s^2_\Theta\tilde G_1^{ij}+
c^2_\Theta\tilde G_2^{ij}\rbrack\Bigr\rbrace\cr
M_{SE}=&(T_{3d}-e_ds_W^2)\sum\limits_{i=1,2}\Bigl\lbrace
V^2_{i1}\lbrack c^2_\Theta\tilde S_{i1}+
s^2_\Theta\tilde S_{i2}\rbrack+
{{m^2_{\rm top}}\over{2m_W^2\sin^2\beta}}
V_{i2}^2\lbrace s^2_\Theta\tilde S_{i1}+
c^2_\Theta \tilde S_{i2}\rbrack\Bigr\rbrace
\cr}$$
The functions $\tilde T^{ab}_i$, $\tilde G^{ij}_a$\
and $\tilde S_{ia}$\ are given in the Appendix B.
After the summation of all diagrams the divergencies
cancel in a nontrivial way. Furthermore in the
case of no mixing $c_\theta=1$\ and  $U_{ij}=
\delta_{ij}=V_{ij}$\ we have $-M_{2\tilde q}+
M_{2\tilde\chi}+M_{SE}\equiv 0$\ after Feynman integration.
Hence with no mixing we would have obtained the same result
as if we had taken
gluinos or neutralinos and scalar down
quarks within the loop; that is,
proportional to the down and strange quark masses
and $q^2$\ and therefore negligible compared to
the SM result.\hfill\break\indent
To obtain the amplitude of the decay
$K^+\rightarrow\pi^+\nu\overline \nu$\
we have to multiply eq.9 by the factor
$\displaystyle{{{g_2}\over{2m_Z^2\cos\Theta_W}}}$.
The amplitude of the decay
$K_L^0\rightarrow\mu^+\mu^-$\
differs by an overall minus sign.
These contributions are shown relative to the SM for different combinations of
masses in Figs.4 and 5. We do not include the quark mixing matrix elements
$K_{td}K^\ast_{ts}$\ and
consider only the effects of masses and couplings. We will discuss the effect
of mixing matrix elements later. One can see that for $\tan\beta\sim 1$ the
contributions can be substantial, even for rather large
SUSY {\it masses}. However, when $\tan\beta\ne 1$ the two scales ($\mu$ and
$m_{g_2}$) must be close and $m_S$ must be relatively small in order to
have a significant contribution. This behaviour becomes more pronounced as the
numerical values increase. Except for particular combinations, one has to
conclude that the contribution will be rather small; unless the SUSY
mixing-matrix elements differ vastly from the SM mixing-matrix elements.

\noindent
\line{\bf Higgs contributions\hfil}
\noindent
Finally we present the results of the contribution
when the charged Higgs boson and the up quarks are taken
within the penguin diagrams.
With the couplings of the charged Higgs to the Z boson
and up quarks given in Fig.2 and Fig.8 in [15]
after summation over all
diagrams and Feynman integration striking cancellations occur and
the finite result is given by:
$$iM={{\alpha^2}\over{ 8\sin^4\Theta_W}}
{{m_{\rm top}^2}\over{m_W^4}}\cot^2\beta
K_{td}K^\ast_{ts}{{m_{\rm top}^2}\over{m_{\rm top}^2-m_{H^+}^2}}
\bigl\lbrack 1-{{m_{H^+}^2}\over{m_{\rm top}^2-m_{H^+}^2}}\ln
{{m_{\rm top}^2}\over{m_{H^+}^2}}\bigr\rbrack
\overline v_s\gamma_\mu P_Lu_d\eqno(10)$$
which basically is the last term of eq.(B4) with $m_i^2
\leftrightarrow m_{\rm top}^2$\ and $m_k^2\leftrightarrow
m_{H^+}^2$. This result agrees with eq.(3.22) in [20] up to
a minus sign, which comes from a relative sign of the couplings
of their eq.(2.6) compared with those given in Fig.2 in [15].
\hfill\break
This contribution, we plot in Fig.6. We see that it can be
substantial in the $K^+\to\pi^+\nu\bar\nu$: it can go up to 25\% for small
masses. The $K^0_L\to\mu^+\mu^-$\ amplitude differs from eq.(10) simply by
a minus sign. It can be as large as 40\% of the short-distance SM contribution.
\hfill\break
\line{\bf Discussion\hfil}
\noindent
In all our previous results, we have used $m_{top} = 174~GeV$ and $\alpha =
1/137$. We have not included any mixing matrix element in our figures.
It is important to remember that $K_{td}K^\ast_{ts}$\
in eq.(8) and eq.(9) have not necessarily the same values as in
the SM. This was shown in eq.(33) in [19]: the
Kobayashi--Maskawa matrix in the couplings of the
charginos to quarks and scalar quarks is multiplied by
another matrix V, which can be parametrized as follows:
$$V=\left(\matrix{1&\varepsilon&\varepsilon^2\cr
 -\varepsilon&1&\varepsilon\cr-\varepsilon^2&-\varepsilon&1\cr}\right)
\eqno (11)$$
\noindent
so that $K\equiv V\cdot K_{SM}$. Clearly,
if $\epsilon\ll 1$ then $K\sim K_{SM}$.
However, with
$\varepsilon=0.5$\ $K_{td}K^\ast_{ts}$\ is enhanced
by a factor of $209$ over the SM value. This enhancement falls quickly as
$\varepsilon$\ decreases: to $28$\ for $\varepsilon=0.3$\
and to $-0.8$\ for $\varepsilon=0.1$. Note that because of the uncertainties
in $K_{SM}$, this last factor has a large error.
We can use that enhancement to try to put limits on $\varepsilon$. For example,
in the $K^+\to\pi^+\nu\bar\nu$ decay,
considering that the amplitude has to be squared,
we can put an upper limit of approximately $0.5$\
if we want the SUSY-penguin contribution to be smaller
than the current experimental limit of approximately 10 times the
SM contribution. Obviously, this is not very constraining but future data might
put interesting contraints on $\varepsilon$ because the penguin contribution to
that process does not vary much with the different SUSY parameters once they
have reached {\it typical} values. Certainly, one would have to know the Higgs
contribution beforehand.

Since the $K_L^0\to\mu^+\mu^-$ decay proceeds mainly via the $\gamma\gamma$
channel [21], we can only say that one has to be carefull in trying to extract
KM matrix elements from that decay: the charged Higgs contribtion has the same
KM elements as the short-distance SM contribution and can reduce the
amplitude by as much
as 40\%. On the other hand, the penguin-chargino contributions remain small
except for some small parts of phase space,
and the box-chargino contributions are negligible.

\noindent\vskip.2cm
\noindent
{\bf IV. CONCLUSIONS}\vskip.2cm
In this paper we presented the results of the calculation of
the 1 loop correction to the decays
$K^+\rightarrow\pi^+\nu\overline \nu$\ and
$K_L^0\rightarrow\mu^+\mu^-$\ within the MSSM.
We gave a complete analysis and included the
mixing of the charginos as well as the mixing of the
scalar partners of the left and right handed top quark.
We have shown that the box diagram contributions are negligible
within the MSSM while the penguin diagram contributions are typically a few
percent of the SM contributions but can be up to 15-20\% for some particular
combinations of $\mu$, $m_{g_2}$, and $m_S$ when $tan\beta\sim 1$
A precise measurement of the process $K^+\to\pi^+\nu\bar\nu$ could lead to a
constraint on the SUSY mixing matrix (the $\varepsilon$ parameter) once
the charged Higgs contribution is known or a bound on its mass is obtained.

In both processes, the charged Higgs contributions to the amplitude
can be quite large
compared to the short-distance contribution from the SM: up to 25\% for the
$K^+\to\pi^+\nu\bar\nu$ and up to 40\% for the $K_L^0\to\mu^+\mu^-$.
Once the top quark mass is well measured, the first decay will open the door
for the observation of the Higgs contribution or constraints on its mass.

\hfill\break\vskip.2cm\noindent
{\bf V. ACKNOWLEDGMENTS}\vskip.2cm
One of us (H.K.) would like to thank the physics department
of Carleton university
for the use of their computer
facilities.
The figures were done with the very user
friendly program PLOTDATA from TRIUMF.
\hfill\break\indent
This work was partially funded by funds from the N.S.E.R.C. of
Canada and les Fonds F.C.A.R. du Qu\'ebec.
\hfill\break\vskip.2cm\noindent
{\bf VI. APPENDIX A}\vskip.2cm
For the box diagram we have to calculate the following
integrals:
$$\eqalignno{F^{\mu\nu}_{abij}:=&\int{{d^4k}\over{(2\pi)^4}}
{{k^\mu k^\nu}\over{(k^2-m_a^2)(k^2-m_b^2)(k^2-m_i^2)(k^2-m_j^2)}}
&(A0)\cr
F^{\mu\nu}_{abij}=:&{{+ig^{\mu\nu}}\over{4(4\pi)^2}}
\tilde F^{ij}_{ab}\cr}$$
$m^2_i\not=m^2_j\not=m^2_a\not=m^2_b$
$$\eqalignno{\tilde F^{ij}_{ab}=-{1\over{(m_j^2-m_i^2)(m_b^2-
m_a^2)}}\bigl\lbrace&{1\over{(m_i^2-m_a^2)(m_i^2-m_b^2)}}
\lbrack m_i^4(m_b^2\ln{{m_b^2}\over{m_i^2}}-m_a^2\ln {{m_a^2}
\over{m_i^2}})\cr
&-m^2_im^2_am_b^2\ln{{m_b^2}\over{m_a^2}}\rbrack-
(m^2_i\leftrightarrow m^2_j)\bigr\rbrace&(A1)\cr}$$
$$\eqalignno{m_j^2\not=&m_i^2\not=m^2_a=m^2_b\cr
\tilde F^{ij}_{aa}=&-{1\over{(m_j^2-m_i^2)}}\bigl\lbrace
{1\over{(m_i^2-m_a^2)}}\lbrack m^2_a-{{m_i^4}\over{(m_i^2-
m_a^2)}}\ln{{m_i^2}\over{m_a^2}}\rbrack-(m_i^2\leftrightarrow
m_j^2)\bigr\rbrace&(A2)\cr
m_i^2=&m_j^2\not=m_a^2\not=m^2_b\cr
\tilde F^{ii}_{ab}=&\tilde F^{ij}_{aa}(m_a^2\leftrightarrow
m_i^2,\ m_b^2\leftrightarrow m_j^2)&(A3)\cr
m_j^2=&m_i^2\not= m_a^2=m_b^2\cr
\tilde F^{ii}_{aa}=&-{{(m_i^2+m_a^2)}\over{(m_i^2-m_a^2)^2}}
\lbrace 1-{{2m_i^2m_a^2}\over{(m_i^4-m_a^4)}}\ln{{
m_i^2}\over{m_a^2}}\rbrace &(A4)\cr}$$
The second integral is given by:
$$\eqalignno{_MF^{ij}_{ab}:=&\int{{d^4k}\over{(2\pi)^4}}
{{m_im_j}\over{(k^2-m_a^2)(k^2-m_b^2)(k^2-m_i^2)(k^2-m_j^2)}}
&(A5)\cr
_MF^{ij}_{ab}=:&{{-ig^{\mu\nu}}\over{(4\pi)^2}} _M\tilde F^{ij}
_{ab}\cr}$$
$m^2_i\not=m^2_j\not=m^2_a\not=m^2_b$
$$\eqalignno{_M\tilde F^{ij}_{ab}=-{{m_im_j}\over{(m_j^2-m_i^2)(m_b^2-
m_a^2)}}\bigl\lbrace&{1\over{(m_i^2-m_a^2)(m_i^2-m_b^2)}}
\lbrack m_i^2(m_b^2\ln{{m_b^2}\over{m_i^2}}-m_a^2\ln {{m_a^2}
\over{m_i^2}})\cr
&-m^2_am_b^2\ln{{m_b^2}\over{m_a^2}}\rbrack-
(m^2_i\leftrightarrow m^2_j)\bigr\rbrace&(A6)\cr}$$
$$\eqalignno{m_j^2\not=&m_i^2\not=m^2_a=m^2_b\cr
_M\tilde F^{ij}_{aa}=&-{{m_im_j}\over
{(m_j^2-m_i^2)}}\bigl\lbrace
{1\over{(m_i^2-m_a^2)}}\lbrack 1-{{m_i^2}\over{(m_i^2-
m_a^2)}}\ln{{m_i^2}\over{m_a^2}}\rbrack-(m_i^2\leftrightarrow
m_j^2)\bigr\rbrace&(A7)\cr
m_i^2=&m_j^2\not=m_a^2\not=m^2_b\cr
_M\tilde F^{ii}_{ab}=&_M\tilde F^{ij}_{aa}(m_a^2\leftrightarrow
m_i^2,\ m_b^2\leftrightarrow m_j^2,\ m_im_j
\rightarrow m_i^2)&(A8)\cr
m_j^2=&m_i^2\not= m_a^2=m_b^2\cr
_M\tilde F^{ii}_{aa}=&-{{m_i^2}\over{(m_i^2-m_a^2)^2}}
\lbrace 2+{{(m_i^2+m_a^2)}\over{(m_i^2-m_a^2)}}\ln{{
m_a^2}\over{m_i^2}}\rbrace&(A9)\cr}$$
\hfill\break\vskip.12cm\noindent
{\bf VII. APPENDIX B}\vskip.12cm
For the penguin diagram we have the following integrals:
$$\eqalignno{\tilde T^{kl}_i=&\int\limits_0^1d\alpha_1
\int\limits_0^{1-\alpha_1}d\alpha_2\bigl\lbrace
{1\over\epsilon}-\gamma+
\ln 4\pi\mu^2-\cr
&\ln\lbrack
m_i^2-(m_i^2-m_k^2)\alpha_1-(m_i^2-m_l^2)\alpha_2
\rbrack\bigr\rbrace&(B1)\cr
=&{1\over 2}\bigl\lbrace {1\over\epsilon}-\gamma
+\ln 4\pi\mu^2-\ln m_i^2\bigr\rbrace+{3\over 4}-
{1\over 2}{{m_l^4}\over{(m_i^2-m_l^2)(m_k^2-m_l^2)}}
\ln{{m_l^2}\over{m_i^2}}\cr
&-{1\over 2}{{m_k^4}\over{(m_i^2-m_k^2)(m_l^2-m_k^2)}}
\ln{{m_k^2}\over{m_i^2}}\rbrace\cr
m_k^2=&m_l^2\not=m_i^2\cr
\tilde T_i^{kk}=&{1\over 2}\bigl\lbrace {1\over\epsilon}-\gamma
+\ln 4\pi\mu^2-\ln m_i^2\bigr\rbrace+{3\over 4}+
{1\over 2}{{m_k^2}\over{(m_i^2-m_k^2)}}\cr
&-{1\over 2}{{m_k^2(m_k^2-2m_i^2)}\over{(m_i^2-m_k^2)^2}}
\ln{{m_k^2}\over{m_i^2}}&(B2)\cr
\tilde G_k^{ij}=&\int\limits_0^1d\alpha_1
\int\limits_0^{1-\alpha_1}d\alpha_2\bigl\lbrace
\bigl\lbrack{1\over\epsilon}-\gamma-1+\ln 4\pi\mu^2\cr
&-\ln\lbrack m_k^2-(m_k^2-m_i^2)\alpha_1-
(m_k^2-m_j^2)\alpha_2\rbrack\bigr\rbrack O^{'L}_{ij}+\cr
&{{m_im_j}\over{\lbrack m_k^2-(m_k^2-m_i^2)\alpha_1-
(m_k^2-m_j)\alpha_2\rbrack}} O^{'R}_{ij}\bigr\rbrace
&(B3)\cr
=&\lbrack \tilde T^{kl}_i(m_k^2\leftrightarrow m_i^2,\
m_l^2\leftrightarrow m_j^2)-{1\over 2}\rbrack O^{'L}_{ij}\cr
&+m_im_j \bigl\lbrack
{{m_i^2}\over{(m_i^2-m_j^2)(m_i^2-m_k^2)}}\ln{{m_i^2}
\over{m_k^2}}+
{{m_j^2}\over{(m_j^2-m_i^2)(m_j^2-m_k^2)}}\ln{{m_j^2}
\over{m_k^2}}\bigr\rbrack O^{'R}_{ij}\cr
m_i^2=&m_j^2\not=m_k^2\cr
\tilde G^{ii}_k=&\lbrack \tilde T^{kk}_i(m_k^2\leftrightarrow
m_i^2)-{1\over 2}\rbrack O^{'L}_{ii}+
{{m_i^2}\over{(m_i^2-m_k^2)}}\lbrack 1-{{m_k^2}\over{
(m_i^2-m_k^2)}}\ln{{m_i^2}\over{m_k^2}}\rbrack O^{'R}_{ii}
&(B4)\cr
\tilde S_{ik}=&\int\limits_0^1\alpha_1\lbrace {1\over\epsilon}
-\gamma+\ln 4\pi\mu^2-\ln\lbrack m_i^2-(m_i^2-m_k^2)
\alpha_1\rbrack
\rbrace&(B5)\cr
=&{1\over 2}\lbrack {1\over\epsilon}-\gamma+\ln 4\pi\mu^2-\ln m_i^2
\rbrack+{1\over 4}+{1\over 2}{{m_i^2}\over{(m_i^2-m_k^2)}}
-{1\over 2}(1-{{m_i^4}\over{(m_i^2-m_k^2)^2}})\ln{{m_k^2}
\over{m_i^2}}\cr
O^{'L}_{ij}=&\cos^2\Theta_W\delta_{ij}-{1\over 2}V_{i2}V_{j2}\cr
O^{'R}_{ij}=&\cos^2\Theta_W\delta_{ij}-{1\over 2}U_{i2}U_{j2}
\cr}$$
\hfill\break\vskip.2cm\noindent
{\bf REFERENCES}\vskip.2cm
\item{[\ 1]} See for example, W.J. Marciano, {\it Kaon Decays} in {\bf Rare
Decay Symposium}, Vancouver, 1988 and references therein; G. B\'elanger, {\it
Rare Decays as Probe for New Physics}, in {\bf Beyond the Standard Model III},
Ottawa, 1992 and references therein; D. Bryman, {\it Rare Kaon Decays and CP
Violation}, in {\bf Fifth Internation Symposium on Heavy Flavour Physics},
Montr\'eal, 1993 and references therein.
\item{[\ 2]} Review of Particle Properties, part 1,
Phys. Rev. {\bf D50}(1994)1172.
\item{[\ 3]} M.K. Gaillard and B.W. Lee, Phys. Rev.{\bf D10}
(1974)897.
\item{[\ 4]} B.W. Lee and R.E. Shrock, Phys. Rev. {\bf D16}
(1987)1444.
\item{[\ 5]}T.Inami and C.S. Lim, Progr. Theor. Phys.{\bf 65}
(1981)297.
\item{[\ 6]}J. Ellis and J.S. Hagelin, Nucl. Phys.
{\bf B217}(1983)189.
\item{[\ 7]}F.J Hilman and J.S. Hagelin, Phys. Lett.
{\bf B133}(1983)443.
\item{[\ 8]}H.E. Haber and G.L. Kane, Phys.Rep.{\bf 117}(1985)75.
\item{[\ 9]}J. Ellis and D.V. Nanopoulos, Phys. Lett. {\bf 110B}
(1982)44.
\item{[10]}C.S. Lim and T. Inami. Nucl. Phys. {\bf B207}
(1982)533.
\item{[11]}A.B. Lahanas and D.V. Nanopoulos, Phys. Lett.
{\bf 129B}(1983)461.
\item{[12]}S. Bertolini and A. Masiero, Phys. Lett. {\bf 174B}
(1986)343.
\item{[13]}B. Mukhopadhyaya and A. Raychaudhuri, Phys. Lett.
{\bf 189B}(1987)203.
\item{[14]}The CDF collaboration (F. Abe et al.),
Phys. Rev. Lett. {\bf 73}(1994)225.
\item{[15]}J.F. Gunion and H. Haber, Nucl. Phys. {\bf B272}
(1986)1.
\item{[16]} A. Djouadi, M.Drees and H. K\"onig, Phys.Rev.
{\bf D48}(1993)3081.
\item{[17]}G. Couture, C. Hamzaoui and H. K\"onig,
"Flavour changing top quark decay within the minimal supersymmetric
standard model", hep-ph@xxx.lanl.gov/9410230.
\item{[18]}H. K\"onig, PhD thesis, unpublished.
\item{[19]}M.J. Duncan, Nucl.Phys.{\bf B221}(1983)221.
\item{[20]}A.J. Buras et al, Nucl.Phys.{\bf B337}(1990)284; see also C.Q. Geng
and J.N. Ng, Phys.Rev.{\bf D38}(1988) 2857 and {\bf D41}(1990) 1715.
\item{[21]}P. Ko, Phys.Rev.{\bf D45}(1992)174, and references therein.
\hfill\break\vskip.2cm\noindent
\noindent
{\bf FIGURE CAPTIONS}\vskip.2cm
\item{Fig.1}The allowed range on $\mu$ and $m_{g_2}$ so that the mass of the
charginos will come out larger than 50 GeV. We used $m_W = 80.1$ GeV and
$\tan\beta =$ 1 (solid line), 2 (long dash), 5 (short dash), 10 (dotted line).
The regions above the upper lines and below the lower
lines are allowed.
\item{Fig.2}The box diagrams with scalar up quarks and charginos
within the loop including the mass insertion diagrams.
\item{Fig.3} The penguin diagrams with scalar up quarks
and charginos within the loop. The mass insertion
diagram with the Z and
chargino couplings leads to the same result and therefore
has not to be included. The diagrams with the charged
Higgs are obtained by replacing the charginos by the
top quark and the scalar top quark by the charged Higgs.
\item{Fig.4} The ratio $Amplitude^{SUSY_{penguin}}/Amplitude^{\rm SM}$\
for the decay $K^+\rightarrow\pi^+\nu\overline \nu$\
as a function of the scalar mass $m_S$
for $\mu = 200$ (solid
line), 500 (dashed line) 1000 (dotted line) GeV and for different values of
$tan\beta$. {\bf NB} We do not include quark-mixing matrix elements.
In {\bf (A)}, we have $m_{g_2} = 200$ GeV while $m_{g_2} = 500$ GeV
 in {\bf (B)}.
\item{Fig.5} The same as Fig.4 but for the decay $K^0_L\to\mu^+\mu^-$
\item{Fig.6} The ratio $Amplitude^{SUSY_{Higgs}}/Amplitude^{\rm SM}$
for the decay $K^+\to\pi^+\nu\bar\nu$ (dashed line) and for the
decay $K^0_L\to\mu^+\mu^-$ (solid line) for $tan\beta = 1$. {\bf NB} We do not
include quark-mixing matrix elements.

\hfill\break
\end